# High Current Density and Low Thermal Conductivity of Atomically Thin Semimetallic WTe$_2$

Michal J. Mleczko[1], Runjie (Lily) Xu[1], Kye Okabe[1], Hsueh-Hui Kuo[2], Ian R. Fisher[3], H.-S. Philip Wong[1], Yoshio Nishi[1], Eric Pop[1,4,*]

1. Department of Electrical Engineering, Stanford University, Stanford CA 94305, USA
2. Department of Materials Science and Engineering, Stanford University, Stanford CA 94305, USA
3. Department of Applied Physics, Stanford University, Stanford CA 94305, USA
4. Precourt Institute for Energy, Stanford University, Stanford CA 94305, USA

**Abstract**

Two-dimensional (2D) semimetals beyond graphene have been relatively unexplored in the atomically-thin limit. Here we introduce a facile growth mechanism for semimetallic WTe$_2$ crystals, then fabricate few-layer test structures while carefully avoiding degradation from exposure to air. Low-field electrical measurements of 80 nm to 2 µm long devices allow us to separate intrinsic and contact resistance, revealing metallic response in the thinnest encapsulated and stable WTe$_2$ devices studied to date (3 to 20 layers thick). High-field electrical measurements and electro-thermal modeling demonstrate that ultra-thin WTe$_2$ can carry remarkably high current density (approaching 50 MA/cm$^2$, higher than most common interconnect metals) despite a very low thermal conductivity (of the order ~3 Wm$^{-1}$K$^{-1}$). These results suggest several pathways for air-stable technological viability of this layered semimetal.

[*]**Contact:** epop@stanford.edu

The preceding decade has seen much interest in two-dimensional (2D) nanomaterials, often exhibiting distinct evolution of chemical and physical properties as material thickness is scaled from layered bulk to individual atomic or molecular monolayers.[1-3] While semiconducting 2D materials have received much attention, layered 2D semimetals other than graphene have been relatively underexplored in the atomically thin limit. Materials like β-MoTe$_2$ and WTe$_2$ stabilize as semimetals in a distortion of the octahedral 1T (CdI$_2$ structure) geometry, with in-plane buckled chains formed by pairs of Mo/W atoms dimerizing in intermetallic charge-exchange,[4-6] while van der Waals bonding dominates interlayer interaction. Whereas MoTe$_2$ may be synthesized in both 2H and 1T' polytypes, or reversibly switched between the two as a function of temperature or strain,[7,8] WTe$_2$ has been known since the 1960s to adopt an orthorhombic structure with space group *Pmn*2$_1$ (sometimes called "Td"), irrespective of growth conditions[4,5,6,9,10] or conventional strain,[8] as the heaviest of the Group VI dichalcogenides.

Despite the inaccessibility of a semiconducting phase, semimetallic WTe$_2$ has received renewed attention from the experimental observation of non-saturating magnetoresistance in bulk samples, in excess of 13,000,000% at 60 T.[11] This behavior was attributed to perfect compensation between balanced electron and hole populations at the Fermi surface below 150 K, projected to persist down to individual monolayers.[12,13] Recent studies have also identified WTe$_2$ as a potential contact for 2D semiconductors, with a relatively low workfunction ($\Phi < 4.4$ eV) amongst 2D metals,[14] recently applied in realizing unipolar *n*-type transport in the typically ambipolar semiconductor WSe$_2$.[15] Layer-dependent experiments of any kind are nonetheless limited,[16-19] owing to a lack of geological sources, challenges in precursor purification during bulk crystal growth,[10,11] as well as observed degradation (oxidation) of thin layered tellurides with exposure to ambient oxygen and moisture.[16,20] In particular, Wang and colleagues studied magnetotransport in uncapped flakes down to bilayer thickness,[19] reporting an insulating regime in sub 6-layer samples attributed to oxidation-induced disorder from ambient exposure.

In this work, we first synthesize bulk WTe$_2$ crystals by a facile growth method employing commercially-available molecular powders, then we isolate few-layer flakes in an inert environment (Supplementary Figure S1). Test devices are fabricated in a manner avoiding any open-air exposure of channel regions, ultimately encapsulating devices *in situ* with AlO$_x$ by atomic layer deposition (ALD), as described in the Methods section. This stabilizes ultrathin WTe$_2$ against ambient degradation, evidenced by spectral analysis of vibrational modes and chemical bonding, preserving Ohmic conduction at high current densities. Electrical characterization is performed on (capped and stable) 3 to 20 layer WTe$_2$ devices using the transfer length method (TLM) approach, separating the contributions of intrinsic and extrinsic (contact) resistance, from 80-300 K. High field measurements (up to breakdown) reveal large current densities needed for contact operation, approaching 50 MA/cm$^2$, in the range of relevance for technological applications.[21-23] By comparison with a self-heating model we are also able to estimate the in-plane thermal conductivity of $WTe_2$. This study represents the successful stabilization and electro-thermal characterization of intrinsic $WTe_2$ approaching the ultimate thickness limit, and could facilitate exploration of further fundamental properties, as potential device contacts, spintronic, memory and interconnect applications.

**Results and Discussion**

**Fabrication and Characterization**

Bulk crystals of $WTe_2$ (Figure 1a) were grown directly by Chemical Vapor Transport (CVT) of a commercial molecular powder (American Elements $WTe_2$, 99.5%), with no need for chemical or thermal precursor pre-treatment, using elemental iodine as a transport agent (see Materials and Methods). We achieved a high yield of few-millimeter-sized crystals, exhibiting both ribbon- and platelet-like morphologies with clear evidence of layered structure under mechanical cleavage or Scanning Electron Microscope (SEM) inspection (Figure 1b-c). Electron Microprobe Analysis (EMPA) confirmed a stoichiometry of $WTe_{2.05}$ with negligible levels of metal contaminants throughout bulk samples. We then mechanically exfoliated few-layer $WTe_2$ flakes onto 90 nm $SiO_2$ on $p^{++}$ Si substrates under an inert atmosphere (a nitrogen-purged glovebox; $O_2$ and $H_2O$ below 3 ppm at their highest levels) and initially capped them with a 300 nm film of poly(methyl methacrylate) (PMMA), serving both as protective coating and resist for electron-beam (e-beam) lithography.

Contacts were lithographically defined, developed and metalized with 20 nm Ti / 20 nm Au, such that exposed device contact surfaces saw cleanroom air for less than 5 minutes before transfer into a load-locked e-beam evaporator (base chamber pressure $\sim 10^{-8}$ Torr). To mitigate the possibility of channel oxidation, we performed resist and metal lift-off in another nitrogen glovebox connected to a thermal ALD chamber where, after lift-off, we immediately deposited ~15 nm of amorphous $AlO_x$ *in situ* by alternating trimethylaluminium (TMA) and $H_2O$ pulses at 150 °C. Inspection by optical and atomic force microscopy (AFM) revealed uniform nucleation of this capping dielectric (Figure 2a,b), with identical RMS roughness on flakes and the surrounding oxide (< 0.4 nm). The smoothness of the capping film facilitated layer counting in flakes directly from AFM height profiles, uniformly measured as integer multiples of the interlayer spacing ~0.704 nm[4,5] with an additional ~0.2 nm offset.

ALD-capped $WTe_2$ nanosheets were found to produce a characteristic, layer-dependent Raman response under illumination from a low-power 532 nm wavelength laser (Figure 1d), consistent with previous reports,[16-19] and lacking any features associated with metal-oxide formation on tungsten dichalcogenides.[24,25] The orthorhombic structure of $WTe_2$ results in a richer set of Raman-active vibrational modes than the simple $A_{1g}$ and $E_{2g}$ (cross- and in-plane) pairing in 2H- layered crystals; a total of 5 modes are identified in the range of 100-250 cm$^{-1}$, with bulk values delineated according to the convention of ref. [17], corresponding primarily to vibrations of Te atoms around an expanded unit cell of a W-W dimer. These soften and stiffen to varying degrees as the layer number is reduced below 10, closely matching known theoretical and experimental values.[17-19] An additional mode appears only in our thin 4L sample, as a blue-shift of the bulk $A_1^7$ mode[18] exposes a 130.5 cm$^{-1}$ shoulder peak (marked by an arrow in Figure 1d). This new feature represents the $A_1^8$ mode identified exclusively in this range by ref. [17].

We utilized the encapsulation to prevent ambient oxidation of ultrathin WTe$_2$, and ALD alumina was chosen for its compatibility with standard microfabrication and effectiveness as an oxygen and moisture barrier (also recently applied for environmental stability of few-layer black phosphorous[26, 27]). Grown on devices whose channels had only seen an inert nitrogen atmosphere (Figures 2a,b), 15 nm AlO$_x$ films were found to significantly improve device yield and preserve Ohmic response with no noticeable current degradation after one week (Figure 2c). In comparison, uncapped devices measured immediately after in-air metal lift-off manifested current non-linearity at moderate source-drain biases, and significant decline in performance over several days - even when stored in partially deoxygenated environments (*e.g.* a tabletop drybox). Such degradation is consistent with increased charge trap density from the progressive oxidation of top-most WTe$_2$ layers, which we evaluated by high resolution X-Ray Photoelectron Spectroscopy (XPS) in Figure 2d.

Oxidation of uncapped crystals is evident in the Te 3d spectra of both freshly cleaved and aged multilayers on SiO$_2$/Si, most prominently in the appearance of secondary peaks across the 3d 3/2 and 5/2 energy range matching reference values for Te(IV)-O binding in TeO$_2$.[28, 29] These features increase in intensity with time relative to Te-W bonds, matching a trend recently observed in ref. [16], though are entirely absent on surfaces probed through the AlO$_x$ capping. There, only W bonding is measured even after 7 days of storage. Ancillary evidence is provided in the upward energetic shift of W-4f peaks in uncapped samples, by an average ~0.8 eV relative to capped flakes, and appearance of a high-energy shoulder suggesting a partial WO$_3$ bonding character induced through atmospheric exposure. Layered WO$_3$ is the oxide most readily formed on W dichalcogenide crystals,[24, 25] producing XPS W-4f reference peaks measured an average 2-3 eV higher[30, 31] in binding energy than those of comparatively closely spaced WS$_2$ and WSe$_2$[31-33] used here as analogues for WTe$_2$. Our findings indicate significant chemical degradation of uncapped layers during the ~ 1 hour period of ambient exposure between glovebox-based exfoliation and XPS measurement, despite prior studies observing constant optical contrast for exposed few-layer samples on the order of 1 day.[19] This supports the conjecture of oxidation-induced disorder driving a metal-to-insulator transition in resistivity when the WTe$_2$ thickness is reduced below 6L,[19] a regime avoided through careful encapsulation in all our thinner samples discussed below.

**Low-Field Electrical Transport**

We performed electrical characterization *via* two-terminal and TLM test structures, with channel lengths $L$ from 80 nm to 2 µm, as shown in Figures 2a,b. As expected for carrier-rich semimetallic devices, the Si back-gate had a negligible (< 5%) effect on current modulation (Supplementary Figure S2) and the remainder of electrical measurements were carried out at zero gate bias. Figure 3a shows the linear fits for a TLM test structure, over the 80-300 K temperature range. Plotting the measured resistance normalized by width, $RW = R_S L + 2R_C$, yields a slope $R_S$ as the intrinsic sheet resistance and the intercept $2R_C$ as the total contact resistance ($L$ and $W$ are the length and width of WTe$_2$ channel). Figure 3b presents a summary of TLM-extracted resistivity $\rho$ in the range 0.4–1.4 mΩ cm (at room temperature) for WTe$_2$ devices of different layer thicknesses. Most ultra-thin devices display metallic behavior ($\rho$ increasing with $T$), consistent with prior reports of bulk resistivity for synthetic WTe$_2$.[1, 18, 34] Only the 17L device exhibited monotonic decline in $\rho$ with increasing $T$; however this was one of the most resistive TLM structures probed, thus its temperature-dependent behavior could be more indicative of defect-limited hopping rather than phonon-limited transport (as for the devices with lower $\rho$).

Interestingly, no clear layer dependence of resistivity emerges for the thickness range probed here; this could be due to different crystalline orientations of the devices, as buckled W chains break the 2D symmetry of the layer plane with a preferred directionality.[4-6] This has been noted in scanning tunneling microscopy (STM) measurements of aligned zigzag features on cleaved WTe$_2$ surfaces,[35] and a strong variation on magnetotransport in bulk ribbons with the angle of the applied field.[11] Thickness invariance in this range is also consistent with recent measurements of an effectively 3-dimensional (3D) electronic structure in WTe$_2$,[36, 37] with only moderate Fermi surface anisotropy in 2D layers attributed to increased interlayer coupling from the described lattice distortion. Room-temperature resistivity remains an order of magnitude greater than that of layered band metals in bulk,[1, 9] including most Group V (V, Nb, Ta) disulfides, selenides and tellurides. It nonetheless remains comparable to that of bulk WTe$_2$,[18, 34] unlike the 10-fold or greater increase of $\rho$ in metals like 1T-TaS$_2$ in the few-nanometer thickness regime.[38, 39]

Extracted contact resistances for 20 nm Ti / 20 nm Au leads (Figure 3c) also show no clear dependence on layer number, with mean $R_C$ spanning a range of 500-600 Ω·µm over the 80-300 K temperature range. (Supplementary Figure S2c displays the temperature dependence of $R_C$, which increases with temperature like the resistivity.) The contact resistance to ultra-thin WTe$_2$ found here is near the lower end of reported resistances for evaporated metals on transition metal dichalcogenides (0.5–2 kΩ·µm) without chemical doping or lattice-level modification.[40, 41]

Figure 3d presents the current density $J$ at 80 and 300 K for all measured thicknesses,[42] at low intrinsic field (4000 V/cm; equivalent to 1 V across a 2.5 µm channel length). Despite the aforementioned variability obscuring an explicit layer dependence, the current density $J$ appears to increase up to 9-11L device thickness, beyond which $J$ saturates. A gradual fall-off in current density might be expected for thicker, metallic layered crystals with top contacts (*i.e.* tens of layers to bulk) due to interlayer resistance limiting current flow to the top-most layers. In contrast, the cross-plane current distribution in few-layer graphene or semiconducting 2D materials is determined by competing effects from electrostatic gating, top-down charge-injection, and interlayer electrostatic screening.[43, 44] For a carrier-rich (semi-) metallic 2D crystal, dielectric screening limits the charge injected into lower layers from top contacts as the thickness is increased, effectively confining current to the upper-most layers in the absence of direct edge-injection. Underlying strata serve primarily to screen-out any substrate (*e.g.* oxide) charge fluctuations in approaching bulk transport limits.[43] Absence of this screening in the thinnest, most sensitive samples (≤ 5L) explains the lower measured current densities and slightly higher TLM-extracted contact resistance.

**High-Field Transport**

We next examine high-field coupled electrical and thermal transport in our WTe$_2$ devices, as summarized in Figure 4. First, we note that for modeling purposes of a given device, $\rho(T)$ can be fit as a function of temperature by a cubic polynomial (Supplementary Figure S2b). This facilitates current calculation as a function of temperature, $I(T) = V/R(T)$, self-consistently with a self-heating (SH) model. To estimate the average temperature rise, we can express the thermal resistance per unit length from the WTe$_2$ channel to the substrate as:[45]

$$g^{-1} = \frac{\mathcal{R}_{\text{Cox}}}{W} + \left\{ \frac{\pi k_{\text{ox}}}{\ln[6(t_{\text{ox}}/W + 1)]} + \frac{k_{\text{ox}}}{t_{\text{ox}}} W \right\}^{-1} + \frac{1}{2k_{\text{Si}}} \left( \frac{L}{W_{\text{eff}}} \right)^{1/2} \quad (1)$$

where $t_{\text{ox}}$ is the SiO$_2$ thickness, $k_{\text{ox}}$ and $k_{\text{Si}}$ are the thermal conductivities of SiO$_2$ and Si (including their temperature dependence[46] see Supplement Section 3) and $W_{\text{eff}}$ is an effective width of the heat dissipation path through the Si substrate.[45] The equation above represents the series combination of three terms: the thermal resistance of the WTe$_2$–SiO$_2$ interface $\mathcal{R}_{\text{Cox}}$, the spreading thermal resistance into the SiO$_2$,[47] and the spreading thermal resistance into the Si substrate (Figure 4a).

The average temperature rise due to Joule heating is $T = T_0 + I(T)V\mathcal{R}_{th}$, where $T_0$ is the ambient temperature and $\mathcal{R}_{th} \approx 1/(gL)$ is the total thermal resistance for "long" devices,[48] much longer than the thermal healing length $L_H$ along our WTe$_2$ devices. Here $L_H = (kWt/g)^{1/2} \approx 70$ to $150$ nm (as we will see in the following discussion), where $t$ is the thickness and $k$ is the lateral thermal conductivity of WTe$_2$. Figure 4c shows that this model with SH can correctly reproduce the decrease in current at high field, whereas the model without SH cannot capture this behavior, for a "long" device with $L \approx 750$ nm. The WTe$_2$–SiO$_2$ thermal interface resistance was used as a fitting parameter here, yielding an estimated $\mathcal{R}_{Cox} \approx 3\times 10^{-8}$ m$^2$ KW$^{-1}$, which is similar to the values of $\mathcal{R}_{Cox}$ for graphene–SiO$_2$ interfaces.[49]

We can also extend this simple SH model to include heat loss from WTe$_2$ to the AlO$_x$ capping layer and to the Ti/Au contacts. This is primarily applicable to our "shorter" devices (compared to $L_H$) where more heat flows laterally into the metal contacts. We can express the peak (maximum) temperature along the WTe$_2$ device as a function of the input power $P$ and other thermal parameters as:

$$T_{max} = T_0 + P\left(\frac{1}{gL}\right)\left(\frac{1 + gL_H\mathcal{R}_T x - 1/\cosh[L/(2L_H)]}{1 + gL_H\mathcal{R}_T x}\right) \qquad (2)$$

where $x = \tanh[L/(2L_H)]$. Similarly, we can also express the average temperature ($T_{avg}$) along the WTe$_2$:

$$T_{avg} = T_0 + P\left(\frac{1}{gL}\right)\left(\frac{1 + gL_H\mathcal{R}_T x - 2xL_H/L}{1 + gL_H\mathcal{R}_T x}\right) \qquad (3)$$

where $\mathcal{R}_T = L_{HM}/[k_m t_m(W+2L_{HM})]$ is the thermal resistance of the metal contacts, $L_{HM} = (t_m t_{ox} k_m/k_{ox})^{1/2}$ represents the thermal healing length into metal contacts of thickness $t_m$ and thermal conductivity $k_m$, and $T_0 = 80$ K or $300$ K. Equations 2 and 3 above reduce to that of the "long" device [$T_{max} \approx T_{avg} \approx T_0 + P/(gL)$] when $L \gg L_T$ and the temperature profile is flat from source to drain.[50] The expressions can also be simplified when contacts are assumed to be perfect heat sinks ($\mathcal{R}_T = 0$), which is often a reasonable approximation.[50] The analytic model given by Equations 1-3 is applicable to most metallic interconnects, not just to WTe$_2$, and it is validated here with finite-element (COMSOL) simulations of the device structure in Figure 4b. (More simulation results are shown in Supplementary Section 5.)

Figure 4d displays several measured $I$–$V$ curves up to breakdown of our WTe$_2$ devices, with high lateral $V_{DS}$ (the higher resistivity at 80 K is due to inter-sample variability of particular 5L and 10L flakes). We find that AlO$_x$-capped WTe$_2$ devices can reach up to 30–50 MA/cm$^2$ current densities, in excess of the 10-20 MA/cm$^2$ benchmark for VLSI interconnect stress-testing.[21-23] We also obtain current densities >30 MA/cm$^2$ in two WTe$_2$ nanoribbons (~50 nm wide) shown in Supplementary Figure S7.

These current densities are larger than typical values for Al and Cu, which are several MA/cm$^2$, and are similar to bulk W films at several tens of MA/cm$^2$.[51-53] Among atomically thin semimetallic layers only graphene can withstand higher current densities, typically hundreds of MA/cm$^2$ and approaching 1 GA/cm$^2$ for aggressively scaled nanoribbons.[45] SEM imaging of most failed devices (Supplementary Figure S6) showed breakdown near the device mid-points, with intact metal contacts, suggesting WTe$_2$ failure at the point of maximum temperature and good contact resistance up to high bias.

Our thinnest AlO$_x$-capped devices (3-5L) in Figure 4d show Ohmic response and breakdown current densities (up to ~50 MA/cm$^2$) comparable to thicker ones. Taken together with Figure 3c-d, these findings are in contrast with a prior study on *uncapped* samples,[19] which reported a sharp increase in resistivity and insulating behavior in ultra-thin WTe$_2$ (< 6L). As another reference point, layered metallic TaSe$_2$ supports lower peak current densities of 19 ± 8 MA/cm$^2$ in conventionally fabricated (also uncapped) devices of ~12 nm thickness, with unreliable measurements in much thinner flakes.[54] Our experiments thus reinforce the importance of encapsulation with AlO$_x$ and avoiding exposure to oxygen and moisture during processing (see Methods). In addition, our simulations (Supplementary Figures S4 and S5) also suggest that the encapsulation layer partly aids lateral heat spreading to the contacts during high-field transport, assisting the higher current densities. Encapsulating WTe$_2$ devices with a higher thermal conductivity material (h-BN instead of AlO$_x$) or placing them on a better heat-sinking substrate (e.g. 30 nm SiO$_2$) could further increase the current densities by another 10-25% (Supplementary Table S2).

**Thermal Conductivity Estimate**

We can also utilize these self-heating studies at high field to estimate the lateral thermal conductivity $k$ of WTe$_2$, following the work of Liao *et al.*[47] with our updated model from equations 1-3 above. The input power is $P = I_D(V_{DS} - 2I_DR_C)$, where $R_C$ is the electrical contact resistance, and $T_{max} \approx 1300$ K is the WTe$_2$ breakdown temperature (the melting temperature of WTe$_2$).[55] For devices capped by AlO$_x$, we must be careful to account for partial lateral heat sinking through this capping layer. Thus, we modify the lateral healing length to $L_H = (k_{eff}Wt/g)^{1/2}$ where the effective thermal conductivity $k_{eff}$ is the parallel combination of lateral heat flow along the WTe$_2$ and the AlO$_x$ capping ($t_{cap} \approx 15$ nm and $k_{cap} \approx 4$ Wm$^{-1}$K$^{-1}$ at high temperature near $T_{max}$).[56, 57] Once $k_{eff}$ is estimated from our SH model, the thermal conductivity of WTe$_2$ can be deduced from $k = k_{eff} - k_{cap}(t_{cap}/t)$.

We note that in this high-temperature breakdown model we cannot fit the thermal conductivity $k$ and $\mathcal{R}_{Cox}$ independently; nonetheless, values consistent with all our measured device breakdowns are fit at $k = $ 2.5-3.5 Wm$^{-1}$K$^{-1}$ for $\mathcal{R}_{Cox}$ of $5 \times 10^{-9}$ m$^2$KW$^{-1}$, up to $k = 9$-11 Wm$^{-1}$K$^{-1}$ for $\mathcal{R}_{Cox} = 10^{-8}$ m$^2$KW$^{-1}$ (see discussion in Supplement Section 3). $\mathcal{R}_{Cox}$ values are expected to be smaller at high temperatures (near $T_{max} \approx 1300$ K) than the earlier $3\times10^{-8}$ m$^2$KW$^{-1}$ estimate at 80-150 K, due to higher phonon occupation.

The lateral thermal conductivity estimated here is greater than that measured by Jana *et al.* on bulk polycrystalline samples of WTe$_2$ (~1 Wm$^{-1}$K$^{-1}$),[18] suggesting higher material quality in exfoliated mono-crystalline flakes, within the range computed by Liu *et al.*[58] The electronic contribution is 10 to 30% of the overall thermal conductivity, based on estimates with the Wiedemann-Franz Law (Supplementary Figure S3). The lower bound of our estimated $k$ is less than half the maximum lattice conductivity of ~9 Wm$^{-1}$K$^{-1}$ along the [100] (in-plane) WTe$_2$ direction, from first-principles calculations.[58] Its magnitude and variation between devices is nonetheless consistent with the anisotropy expected between multiple in-plane (*i.e.* relative to W-W dimer chain orientation) and cross-plane $k$ values, suggesting a strong role of structural asymmetry on thermal transport in such crystals. Orientation-mapping of devices and ancillary measurement techniques (*e.g.* time-dependent thermoreflectance across flakes of varying thickness) are needed to elucidate the directional-dependence of this parameter.[59]

**Conclusion**

In conclusion, we studied electrical and thermal transport in ultra-thin (3-20L) semimetallic WTe$_2$ devices. The WTe$_2$ crystals were grown from a commercial molecular powder and exfoliated as few-layer flakes in an inert atmosphere, on which we fabricated TLM test structures. Glovebox-based processing and *in situ* encapsulation with an ALD alumina layer protected devices from ambient oxidation, this process being essential for obtaining good transport and stability across several weeks for the thinnest (3-6L) devices. The intrinsic resistivity of our ultra-thin WTe$_2$ is $10^{-4}$–$10^{-3}$ Ω·cm between 80–300 K, with mean contact resistances of 400–600 Ω·μm. The maximum current density ranged from 30-50 MA/cm$^2$ in encapsulated, air-stable devices (including WTe$_2$ nanoribbons), which is higher than that achievable in most bulk metal interconnects. Comparison of high-field breakdown with an analytical self-heating model estimated low intrinsic thermal conductivity around 3 Wm$^{-1}$K$^{-1}$ for such ultra-thin WTe$_2$ devices. Additional finite-element simulations indicate that the maximum current density of these interconnects could be increased by capping with a higher thermal conductivity material (like h-BN) or placing on better heat-sinking substrates (*e.g.* thinner SiO$_2$).

It is tempting to assign thermoelectric applications to a good conductor of electricity with poor thermal properties, however the thermopower (Seebeck coefficient) of WTe$_2$ is relatively small due to the semi-metallic nature.[18, 34] Nevertheless, applications in phase-change memory particularly demand nanoscale electrodes with good current density and poor thermal conductivity,[60, 61] to lower the programming energy per bit. In addition, nanostructured WTe$_2$ could also be a promising candidate for other applications, including as 2D contacts to layered transistors,[14, 15, 60] in magnetic memory,[11, 19] sensors and spintronics.[62, 63]

## Methods

### Material Growth and Fabrication

Bulk $WTe_2$ crystals were grown by Chemical Vapor Transport of a $WTe_2$ molecular powder (American Elements, 99.5%) sealed in a quartz tube evacuated under argon, with elemental iodine (Alfa Asear, 99.99+%) added at 5 mg/cm$^3$. Growth took place for 14 days along an 11 cm transport length, in a single-zone furnace with a central temperature of 900 °C and a ~100 °C thermal gradient. (Supplementary Figure S1.) Few-layer flakes were exfoliated onto 90 nm $SiO_2$ on p$^{++}$ Si substrates within the inert atmosphere of a nitrogen glovebox, using low-residue thermal release tape, and were solvent cleaned and capped *in situ* by spin-coating a layer of PMMA (Microchem A5 950k). Certain substrates were subject to a weak $O_2$ plasma exposure (2 minutes at 60 W, 250 mTorr) to promote adhesion of thinner flakes. The protective PMMA layer also served as a resist for electron-beam lithography of top-contacts (Raith 150, 20 kV), developed and transferred within <5 min. into load-locked metal evaporators (Kurt J. Lesker or AJA, both electron beam) for deposition of 20 nm Ti / 20 nm Au. Liftoff was performed with acetone/2-propanol in a nitrogen glovebox directly connected to a Cambridge Savannah Thermal ALD system. Thus, ALD encapsulation (with ~150 Å of alumina by alternating TMA and water cycles at 150 °C, first saturating surfaces with 10 leading TMA cycles) was accomplished without exposing the devices to ambient air.

### Characterization

Compositional analysis on bulk crystals was performed with a JEOL JXA-8230 Electron Probe Microanalyzer. Individual flakes were profiled with an AFM (Veeco Dimension 3100) and Raman Spectroscopy (Horiba Labram, 532 nm laser source). High resolution XPS analysis was performed in a Phi 5000 VersaProbe, calibrated to surface Carbon 1s peaks, with capping layers etched away by *in situ* Ar$^+$ sputtering with iterative signal collection, such that an Al signal was monitored to prevent damage to the flake-alumina interface. Bulk crystals and devices were imaged at high-resolution with a FEI XL30 Sirion and the SEM mode of a Raith 150. Electrical characterization was performed in a Janis Cryogenic probe station (chamber pressure $10^{-6}$ to $5\times10^{-5}$ Torr) cooled with closed-loop liquid nitrogen and connected to a Keithley 4200-SCS parameter analyzer.

### Acknowledgements

Work was performed at the Stanford Nanofabrication Facility (SNF) and Stanford Nano Shared Facilities (SNSF).We acknowledge fruitful discussions with P. McIntyre, J. Provine, M. Rincon, and J. Conway, and technical assistance from A. Hazeghi with crystal SEM and R. Jones with EMPA analysis. This work was supported in part by the Air Force Office of Scientific Research (AFOSR) grant FA9550-14-1-0251, the National Science Foundation EFRI 2-DARE grant 1542883, and the Initiative for Nanoscale Materials and Processes (INMP). M.J.M. would like to acknowledge an NSERC PGS-D fellowship.

**Supporting Information**

The Supporting Information is available free of charge on the [ACS Publications website](#) at DOI: [to be filled in by ACS]. Figures showing the temperature dependence of 2-terminal resistivity and its gate dependence; elucidation of the self-heating model and thermal conductivity estimates; temperature evolution of extracted contact resistance and the electronic contribution to thermal conductivity; finite-element (COMSOL) simulations of device temperatures for various substrate oxide thicknesses and encapsulation layer configurations; SEM analysis of bulk flakes and laterally-scaled nanoribbons following breakdown; discussion of $WTe_2$ breakdown mechanisms.

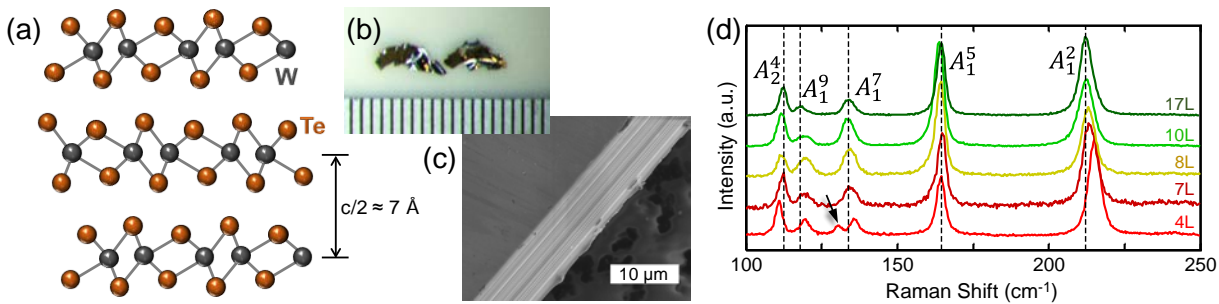

**Figure 1.** (a) Schematic cross-section of semimetallic orthorhombic phase of $WTe_2$.[5, 11, 18] (b) Bulk $WTe_2$ crystals grown by CVT, with mm increments for scale. (c) SEM micrograph of grown bulk crystal displaying layered structure at the edge. (d) Raman spectra of ALD-capped few-layer $WTe_2$, labeling typically observed modes (Horiba Labram, 532 nm laser at 2.5 mW power). The arrow marks the $A_1^8$ mode which appears in our thin 4L sample.

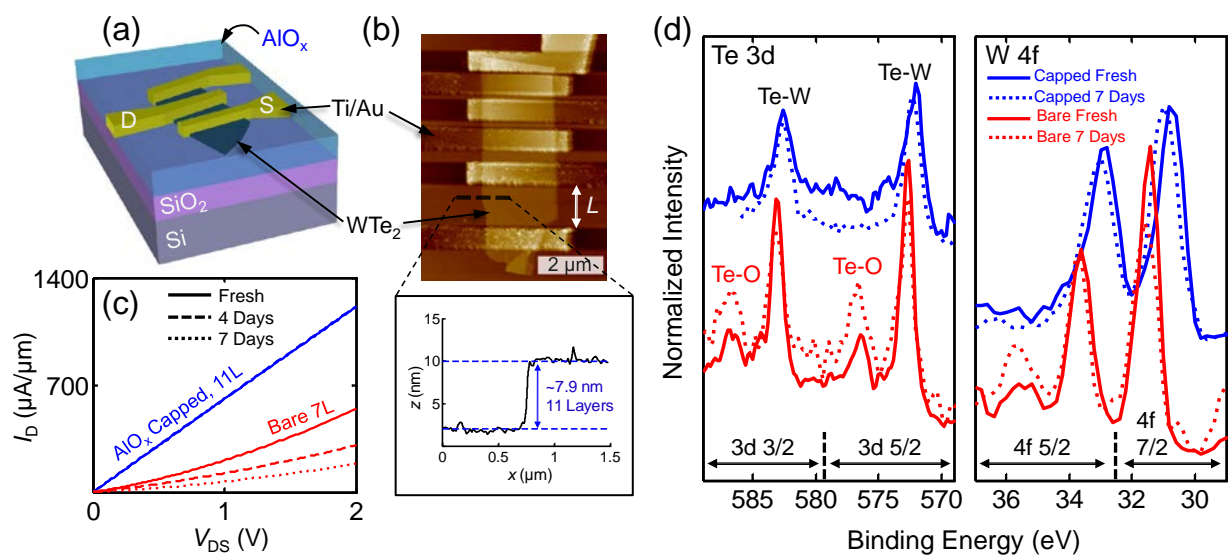

**Figure 2.** (a) Schematic of AlO$_x$-capped WTe$_2$ TLM structures on 90 nm SiO$_2$ (on Si) with Ti/Au contacts. (b) AFM micrograph of fabricated WTe$_2$ TLM structures showing 5 device channel lengths (from 85 to 1500 nm) and 6 electrodes, capped with ~15 nm AlO$_x$ by ALD. AFM height profile (lower inset) was extracted along the dotted line. (c) Time-dependent degradation of current vs. voltage ($I_D$–$V_{DS}$) in AlO$_x$ capped (11 layer, $L$ = 0.5 μm) and uncapped (7 layer, $L$ = 0.46 μm) devices. The ALD-capped devices are air-stable for over one week, whereas the bare (uncapped) devices degrade within hours or days. (d) High-resolution XPS of ALD-capped and bare multilayer WTe$_2$ flakes on SiO$_2$/Si substrates; ambient degradation is visible in the formation of Te-O sub-peaks and a binding energy shift of W-4f peaks suggesting partial WO$_3$ bonding character.

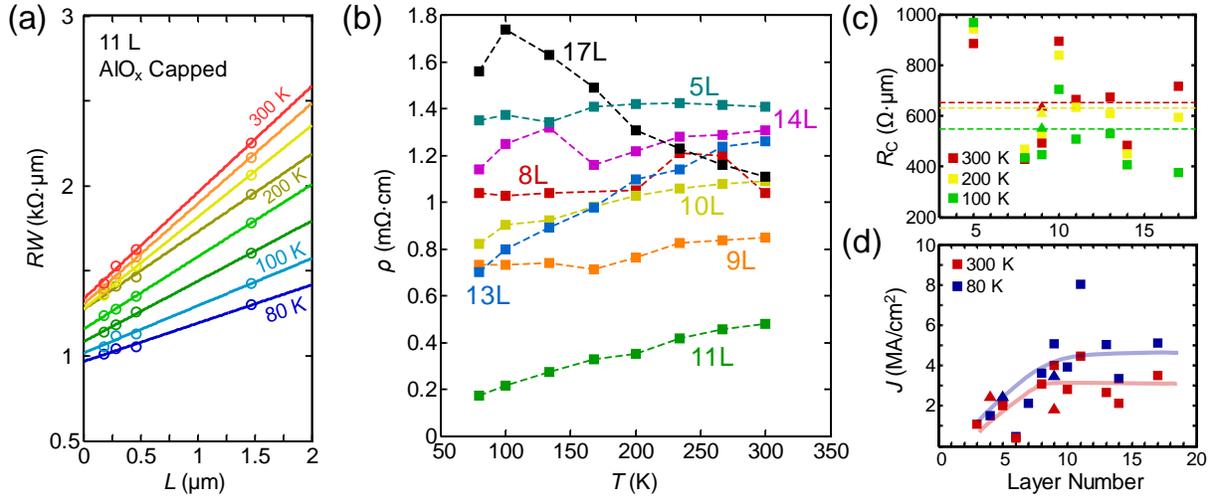

**Figure 3.** (a) Typical TLM plot of AlO$_x$ capped, 11-layer WTe$_2$ devices presenting the total resistance (normalized by width) vs. channel length $L$; lines are a numerical fit to measured values (symbols). Figure 2a-b display the typical TLM geometry. The vertical intercept yields $2R_C$ and the slope yields $R_S$. (b) Measured temperature dependence of resistivity for AlO$_x$ capped WTe$_2$ devices with 5–17 layers, derived from TLM sheet resistance. (c) Contact resistance between WTe$_2$ and 20 nm Ti / 20 nm Au contacts for different temperatures and layer numbers, extracted from TLM measurements. Dashed lines denote average $R_C$ values at 100, 200 and 300 K. (d) Low-field ($F = 4000$ V cm$^{-1}$) current density in capped WTe$_2$ devices at 80 and 300 K, corrected for contact resistance. Lines are guides to the eye; different symbols represent different samples.

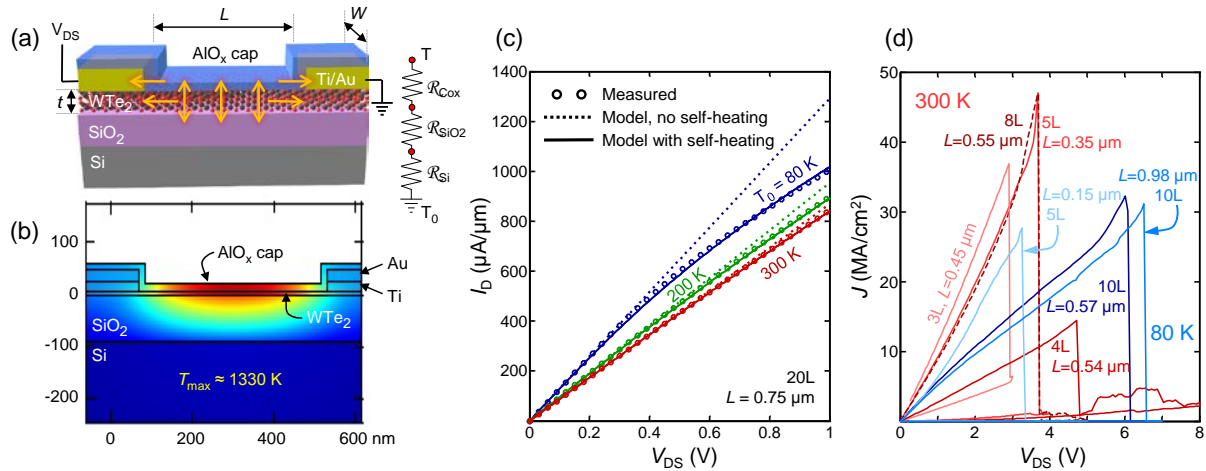

**Figure 4.** (a) WTe$_2$ device schematic, with arrows showing pathways for heat sinking during high-field operation. Inset shows simplified thermal resistance model for vertical heat flow, which dominates in longer devices ($L \gg L_H$). (b) Thermal simulations (COMSOL) were used to validate the analytic thermal model (Supplementary Section 5). Here the simulation is shown just before WTe$_2$ breakdown. (c) Measured (symbols) and simulated (lines) current vs. voltage for an uncapped 20 layer WTe$_2$ device. Solid lines modeled with self-heating (SH) model, dashed lines are without SH. (d) Measured current density vs. voltage up to thermal breakdown of ALD-capped WTe$_2$ devices at 80 K (blue) and 300 K (red) ambient in vacuum probe station (~$10^{-5}$ Torr). The maximum current density approaches 50 MA/cm$^2$, almost an order of magnitude higher than typical bulk metal interconnects (*e.g.* Al, Cu). SEM images after device breakdown are shown in Supplementary Figure S6.

**Table of Contents Figure**

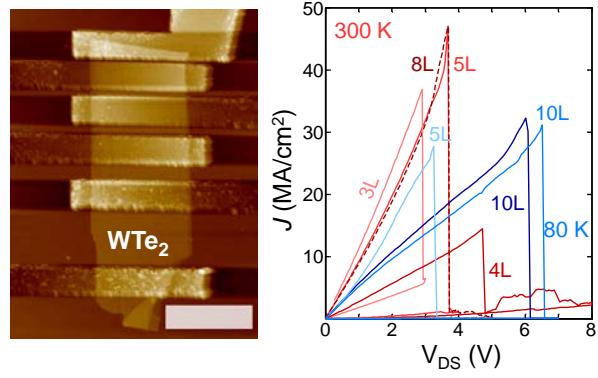

# Supplementary Information

# High Current Density and Low Thermal Conductivity of Atomically Thin Semimetallic WTe$_2$

Michal J. Mleczko[1], Runjie (Lily) Xu[1], Kye Okabe[1], Hsueh-Hui Kuo[2], Ian R. Fisher[3], H.-S. Philip Wong[1], Yoshio Nishi[1], Eric Pop[1,4,*]

1. Department of Electrical Engineering, Stanford University, Stanford CA 94305, USA
2. Department of Materials Science and Engineering, Stanford University, Stanford CA 94305, USA
3. Department of Applied Physics, Stanford University, Stanford CA 94305, USA
4. Precourt Institute for Energy, Stanford University, Stanford CA 94305, USA

# 1. Growth and Exfoliation

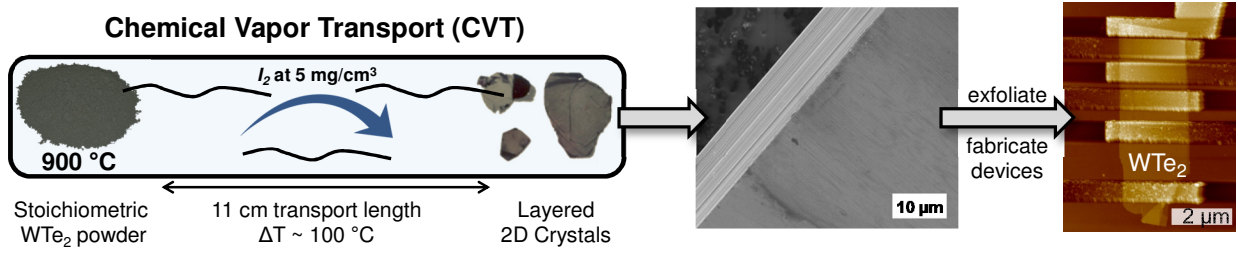

**Figure S1.** Schematic of CVT growth (see Methods) resulting in mm-size crystals, which are exfoliated into ultra-thin flakes (3-20 layers) to produce the devices studied in this work.

# 2. Electrical Characterization

Figure S2a demonstrates the independence of Ohmic behavior in few-layer WTe$_2$ flakes on electrostatic gating, with minute variation in DC current across $V_G$ = -38 to +38 V sweep of a global back-gate (90 nm SiO$_2$ on p$^{++}$ Si). Such invariance was reproduced across all layer thicknesses and temperatures in the 80-300 K range, on both capped and uncapped samples.

Figure S2b demonstrates representative dependence of low-field resistivity vs. temperature ($\rho$ vs. $T$) for both capped and freshly exfoliated, uncapped flakes in the Ohmic regime. A cubic polynomial fit can be made to both curves (dashed lines), as is implemented in the described self-heating model (Figure 4c). Monotonic increase in resistivity with temperature is consistent with metallic transport.

Figure S2c presents the temperature dependence of contact resistance ($R_C$), extracted as one-half of the $y$-intercept in Transfer Length Measurements (TLM) of total resistance vs. channel length (Figure 3a,c). $R_C$ increases monotonically with temperature, but its dependence is less strong than that of resistivity $\rho$ (Figure 3b and S2b). Both $\rho$ and $R_C$ show no clear dependence on layer number. This is consistent with the prior discussions of a relatively 3-dimensional (3D) electron Fermi surface in this structurally 2D material, in addition to the different crystalline orientation of the devices.

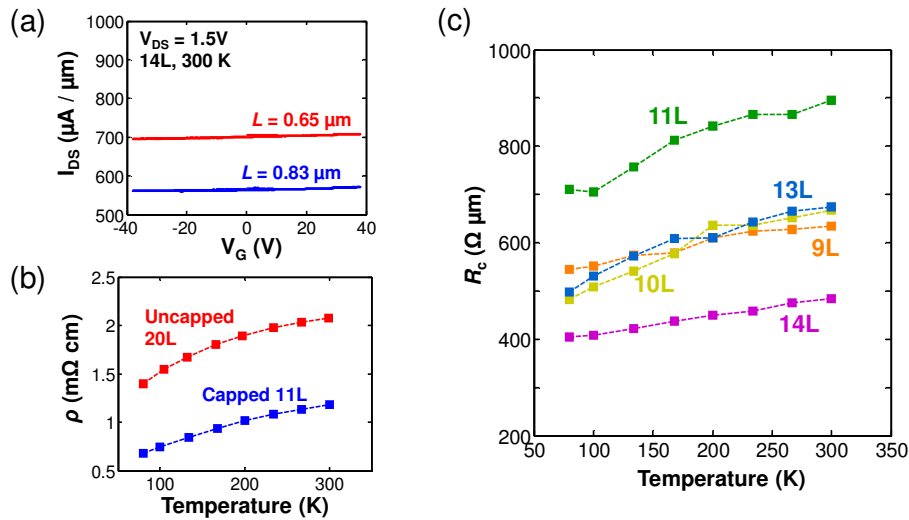

**Figure S2. (a)** Lack of gate voltage dependence for an AlO$_x$ capped 14 layer WTe$_2$ flake on 90 nm SiO$_2$ and p++ Si global back-gate, showing negligible current modulation. **(b)** Comparative temperature-dependent resistivity of newly made capped and uncapped flakes, fit with a cubic polynomial model (dashed lines). **(c)** Temperature dependence of TLM-extracted $R_C$ for AlO$_x$-capped WTe$_2$ multilayers.

## 3. Thermal Conductivity Estimates

A schematic of the heat flow pathways of WTe$_2$ devices on 90 nm SiO$_2$ and Si substrates is provided in Figure 4a of the main text. The average temperature rise of a "long" device ($L \gg L_T$) is $\Delta T = T - T_0 = P\mathcal{R}_{th}$ where $P$ is the input power (i.e. Joule heating), $\mathcal{R}_{th}$ is the overall thermal resistance, and $T_0$ is the ambient temperature. As described in the main text with further details provided below, the thermal resistance $\mathcal{R}_{th}$ has temperature dependence through the thermal conductivity of SiO$_2$ ($k_{ox}$) and the thermal conductivity of the doped silicon substrate ($k_{Si}$). Their evolution with temperature $T$ can be captured analytically as $k_{ox} = \ln(T_{ox}^{0.52}) - 1.687$ and $k_{Si} = 2.4 \times 10^4 / T_0$, where the average temperature of the SiO$_2$ is $T_{ox} = (T + T_0)/2$ and the substrate is assumed at the ambient temperature $T_0$.[1] This enables a simple iterative solution of the average device temperature $T$, obtained self-consistently with $k_{ox}(T)$, $k_{Si}(T)$, and the measured device resistivity $\rho(T)$, which is also a function of the average device temperature.

The current is calculated as a function of temperature, $I(T) = V/R(T)$, self-consistently with the self-heating (SH) model whose input are the power $P = I(T)V$ and thermal resistance $\mathcal{R}_{th}(T)$. The input power may be corrected for the voltage lost at the contacts, i.e. $P = I(V - 2IR_C)$.

Following previous work on Joule heating in graphene nanoribbons (GNRs),[2, 3] we can express the peak (max) temperature along the WTe$_2$ device as a function of input power and other thermal parameters as:

$$T_{max} = T_0 + P\left(\frac{1}{gL}\right)\left(\frac{1 + gL_H R_T x - 1/\cosh[L/(2L_H)]}{1 + gL_H R_T x}\right) \quad (S1)$$

where $x = \tanh[L/(2L_H)]$. Similarly, we can also express the average temperature ($T_{avg}$) along the WTe$_2$:

$$T_{avg} = T_0 + P\left(\frac{1}{gL}\right)\left(\frac{1 + gL_H R_T x - 2x L_H/L}{1 + gL_H R_T x}\right) \quad (S2)$$

In the limit of a "long" device ($L \gg L_H$), the expressions above reduce to $T \approx T_{avg} \approx T_{max} = T_0 + P/(gL)$ and $g$ is described in the main text. If the WTe$_2$ channel width is small (comparable to the thickness of back-gate SiO$_2$ thickness 90 nm), the fringing effect of heat loss to the SiO$_2$ substrate must be considered. We adjust for an effective width $W_{eff} = W + 2t_{ox}$ instead of measured width $W$ to describe this fringing effect (as shown by lateral arrows representing the neat heat diffusion pathway in Figure 4b). In the Equation S1 and S2, $R_T$ is the thermal resistance of the metal contacts, $L_H$ is the thermal healing length along the WTe$_2$, as described in the main text. Comparing this simple model with temperature-dependent $I$-$V$ data (Figure 4a), we can estimate the WTe$_2$-SiO$_2$ contact thermal resistance, $\mathcal{R}_{Cox} \sim 3\times10^{-8}$ m$^2$ KW$^{-1}$ in the temperature range 80–150 K. This simple estimate is possible because for "long" devices the heat sinking occurs almost entirely into the substrate and the contacts play very little role.

In the limit of a "short" device ($L$ comparable to or shorter than $L_H$) heat sinking can occur both into the substrate and into the two contacts. By calculating the lateral heat sinking component we can estimate the lateral thermal conductivity of WTe$_2$. We note that our devices were capped with AlO$_x$, which means that lateral heat flow is given by an effective thermal conductivity, $k_{eff}$, which is the parallel combination of heat flow through the WTe$_2$ and AlO$_x$ capping layer (see Figure 4a). $k_{eff}$ enters the equations above through the lateral thermal healing length, $L_H = (k_{eff}Wt/g)^{1/2}$. The actual value of WTe$_2$ thermal conductivity is deduced as $k = k_{eff} - k_{cap}(t_{cap}/t)$, where $k_{cap} \approx 4$ Wm$^{-1}$K$^{-1}$ (near $T_{max}$) is the thermal conductivity of AlO$_x$ and $t_{cap} \approx 15$ nm its thickness.[4, 5]

As in previous studies of Joule heating in GNRs,[2, 3] we can estimate the lateral thermal conductivity $k$ of WTe$_2$ by taking advantage of electrical $I$-$V$ measurements taken up to device breakdown, reaching $T_{max} \approx$ 1300 K (see Section 4 below). We use only "short" WTe$_2$ devices for this estimate, where lateral heat sinking to the contacts plays a non-negligible role through the thermal conductivity $k$. (As stated earlier, "long" devices sink most of their heat into the substrate.) Using equation S1 above, we simply relate the (maximum) power input at the point of breakdown from the *I-V* measurements in Figure 4d to the thermal conductivity, which enters equation S1 through the healing length $L_H$. We use the shorter three devices measured up to breakdown in Figure 4d to fit $\mathcal{R}_{Cox}$ and $k$ at the same time to match $T_{max} \approx 1300$ K. We note this represents a "high temperature" (near breakdown) thermal conductivity of WTe$_2$.

Values fit in this manner, listed in Table S1, match devices breakdown profiles consistently:

**Table S1.** Lateral thermal conductivity of WTe$_2$ extracted from device breakdown profiles, per values of WTe$_2$-SiO$_2$ thermal contact resistance used as a fitting parameter.

| $\mathcal{R}_{Cox}$ (m$^2$ KW$^{-1}$) | $k$ (Wm$^{-1}$K$^{-1}$) |
|---|---|
| 5×10$^{-9}$ | 2.5 to 3.5 |
| 7.5×10$^{-9}$ | 4 to 6 |
| 10$^{-8}$ | 9 to 11 |

The lowest extracted $k$ values exceed those measured by Jana et al. (~1 Wm$^{-1}$K$^{-1}$) on bulk samples,[6] necessitating a boundary resistance less than one-tenth the $\mathcal{R}_{Cox} = 3 \times 10^{-8}$ m$^2$KW$^{-1}$ estimate at 80-150 K. A decline in $\mathcal{R}_{Cox}$ is expected at elevated temperatures, with higher occupation of relevant phonon modes.

### 4. Electronic Contribution of Thermal Conductivity

The electronic contribution $k_e$ to the net thermal conductivity $k$ of few-layer WTe$_2$ flakes is calculated by the Wiedemann-Franz Law as $k_e = \sigma L T$, where $\sigma$ is electrical conductivity, $T$ is temperature, and $L = 2.44 \times 10^{-8}$ W Ω K$^{-2}$ is the Lorenz number. Estimated $k_e$ values for WTe$_2$ devices of varying thickness are shown in Figure S3, extracted for AlO$_x$-capped few-layer films. This electronic contribution in ultrathin devices is consistent with prior measurements on bulk, polycrystalline samples through a similar Wiedemann-Franz interpretation of crystal resistivity.[6] Combining these observations with our estimates above, we surmise that total thermal conductivity of WTe$_2$ is dominated by phonons, but with a non-negligible (10-30%) electronic contribution.

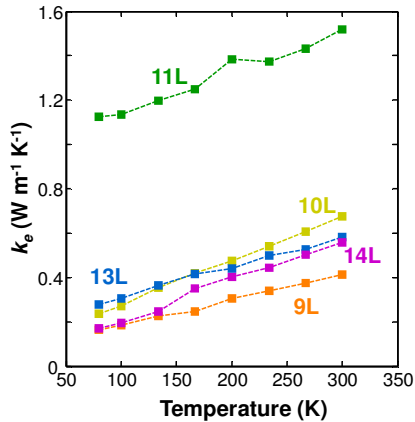

**Figure S3.** Electronic contribution ($k_e$) to the lateral thermal conductivity of WTe$_2$ multilayers vs. temperature. Estimates were made with the Wiedemann-Franz Law from the resistivity directly measured on our TLM structures.

## 5. Finite Element Simulations

We use finite-element (FE) simulations in COMSOL to verify our analytic thermal model, with results shown in Figures S4 and S5. The FE simulations confirm the predictions of our analytic model, and the ranges of $k$ and $\mathcal{R}_{Cox}$ used therein. Furthermore, the FE simulations also indicate that the device temperature can be reduced by two methods: using a high thermal conductivity material (such as $h$-BN) as the capping layer, and decreasing the $SiO_2$ substrate thickness (Figure S4c-d, respectively). $h$-BN capping provides improved heat spreading to the contacts due to its large in-plane thermal conductivity,[7] whereas decreasing the $SiO_2$ thickness reduces the total thermal resistance of the substrate.

Figure S5 directly compares the predictions of the analytic model (dashed lines) and of the COMSOL simulations (symbols). The maximum channel temperature $T_{max}$ vs. applied power $P$ is shown for uncapped, $AlO_x$ capped, and $h$-BN capped $WTe_2$ devices on 90 nm $SiO_2$, and an $AlO_x$ capped device on 30 nm $SiO_2$. Table S2 summarizes how $h$-BN capping and a thinner $SiO_2$ substrate could help $WTe_2$ devices reach higher current densities (> 50 MA/cm$^2$) before breakdown.

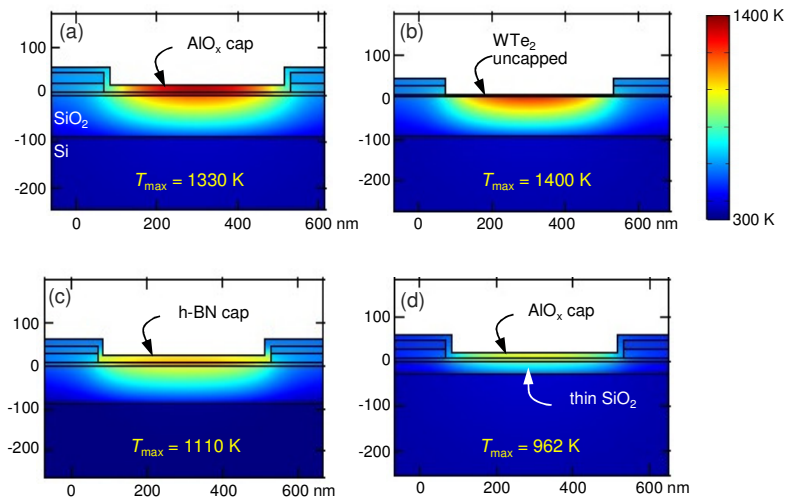

**Figure S4.** Temperature distribution from finite-element (COMSOL) simulations with 3.55 V applied across a $WTe_2$ device (dimensions $L$ = 465 nm, $W$ = 1.1 μm and $t$ = 6.3 nm). Devices are **(a)** capped with 15 nm $AlO_x$ as in the experiments, **(b)** uncapped, and **(c)** hypothetically capped with 15 nm $h$-BN (all on 90 nm $SiO_2$ on an Si substrate). Device **(d)** is capped with 15 nm $AlO_x$ on 30 nm $SiO_2$ substrate. The maximum temperature $T_{max}$ in (a) reaches the melting temperature of $WTe_2$, consistent with experiments in Figure 4d of the main text and the breakdowns shown in Figure S6. All simulations use temperature-dependent thermal conductivities of Si, $SiO_2$ and $h$-BN.

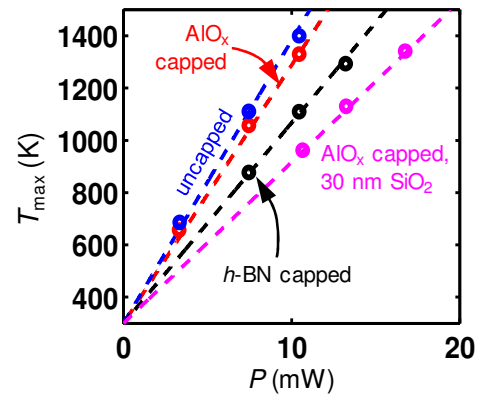

**Figure S5.** Maximum channel temperature ($T_{max}$) vs. power input ($P$) for a $WTe_2$ device (device dimensions as in Figure S4), comparing analytical model (dashed lines) and simulation results (symbols), validating the use of the analytic model. The same four cases from Figure S4 are studied (uncapped device, capped with $AlO_x$ on 90 nm $SiO_2$, capped with $h$-BN, or capped with $AlO_x$ on 30 nm $SiO_2$). We use an effective thermal conductivity[7] for anisotropic $h$-BN in the analytical model, $k_{BN}$ = $(k_{\parallel}k_{\perp})^{1/2}$ ~ 15 W/m/K (at high temperature, near $T_{max}$).

**Table S2.** Summary of FE simulations, using $k = 10$ Wm$^{-1}$K$^{-1}$ and $\mathcal{R}_{Cox} = 10^{-8}$ m$^2$KW$^{-1}$ for WTe$_2$ (device dimensions as in Figure S4), comparing maximum achievable voltages and current densities of $h$-BN capped, AlO$_x$ capped and uncapped devices at $T_{max} \sim 1300$ K (breakdown). The $h$-BN capped device achieves the highest current density ($J_{max}$). Thermal conductivity values are used at high temperature, where data are available, including the anisotropy of $k_\parallel$ and $k_\perp$ in $h$-BN.[5, 7, 8]

| Capping material | $t_{cap}$ (nm) | $t_{SiO2}$ (nm) | $k_{cap}$ (W/m/K) | $V_{max}$ (V) | $J_{max}$ (MA/cm$^2$) |
|---|---|---|---|---|---|
| $h$-BN | 15 | 90 | ~150 (∥, in-plane) <br> ~1.5 (⊥, cross-plane) | 4 | 48 |
| AlO$_x$ | 15 | 90 | ~4 | 3.55 | 43 |
| AlO$_x$ | 15 | 30 | ~4 | 4.5 | 54 |
| none | | | | 3.4 | 41 |

## 6. Thermal Breakdown

Figure S6 presents SEM micrographs of few-layer, AlO$_x$-capped WTe$_2$ devices following high-current breakdown, imaged at 5 and 10 kV with the SEM mode of a Raith 150. Failure occurs near the mid-point of the channel for all samples, and the Ti/Au electrodes remain fully intact in imaged devices (top and bottom regions of the Figure S6 images). In particular, breakdown around channel centers is consistent with measured metallic conduction, failing nearest the point of highest temperature, $T_{max}$. In certain short-channel devices, some asymmetry of the breaking point was noted towards a particular electrode, suggesting local field non-uniformity.

Near-complete rupture is observed across the entire width of flakes, alongside apparent local disruption of the encapsulating dielectric layer in certain devices (i.e. both 10L channels). The failure temperature of WTe$_2$ $T_{max} \approx 1300$ K is nonetheless 300-400 K below the range for onset of melting in AlO$_x$, even for amorphous films with small polycrystalline grain size such as those produced by atomic layer deposition (ALD).[9] Locally defective capping layers may reduce this threshold. Moreover, the WTe$_2$ melting temperature coincides with the onset of several secondary phenomena, including the boiling point of Tellurium (producing buried gas bubbles which escape by rupturing capping layers),[10] and the formation of Al-Te glasses.[11] Volume expansion from the latter reactions may explain the oxide stress visible around failed devices with intact dielectrics (i.e. the 8L device in Figure S6). Our ALD deposition of AlO$_x$ uses H$_2$O precursor, which leads to Al-rich and sub-stoichiometric oxygen within capping layers.[12] A confluence of these factors likely produces the observed state of post-breakdown devices.

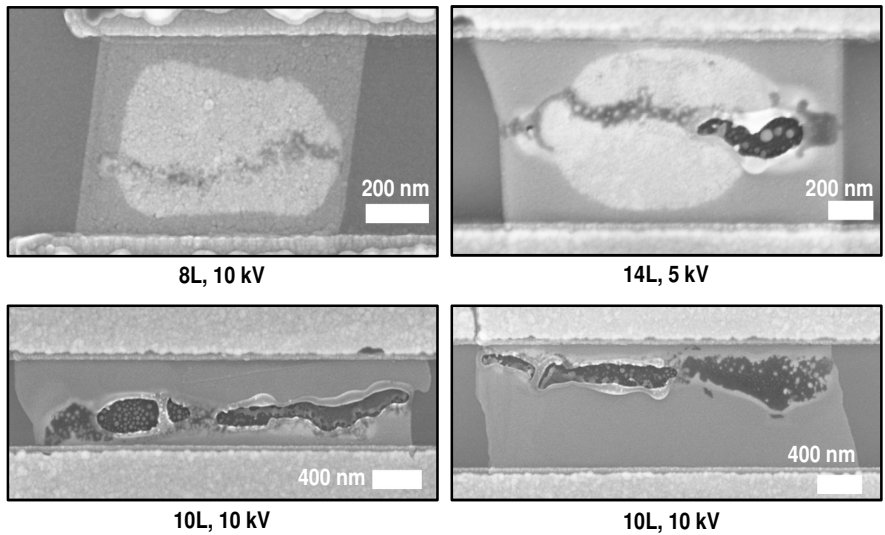

**Figure S6.** SEM micrographs of AlO$_x$-capped WTe$_2$ channels following high-current thermal breakdown, imaged at 5 and 10 kV. Uniform breakdown in channel centers signifies uniform heating during operation and negligible contact resistance. 8 and 10-layer (8L and 10L) samples correspond to devices broken-down in Figure 4d, at respective ambient of $T_0 = 300$ and 80 K.

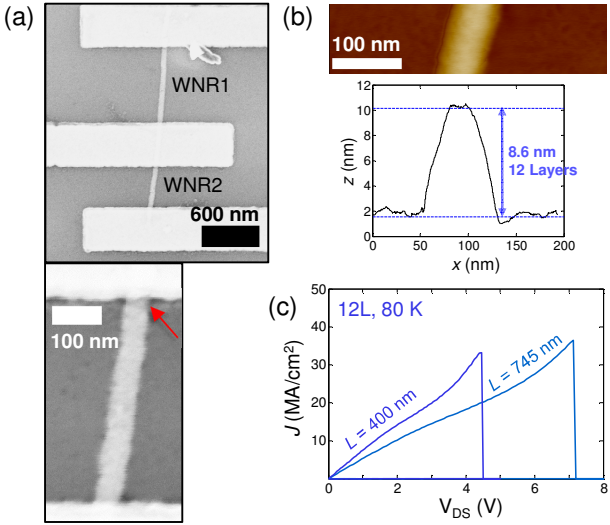

**Figure S7.** (a) SEM micrographs and (b) AFM micrographs and height profile of a ~50 nm wide, 12L thick exfoliated $WTe_2$ nanoribbon (WNR), capped with $AlO_x$. Red arrow indicates point of failure. (c) Breakdown current densities of the two WNRs (at 80 K ambient) are comparable to the large $WTe_2$ device current densities reported in Figure 4d.

Figure S7 shows results of as-exfoliated, 12L $WTe_2$ nanoribbon (WNR) devices, electrically driven to the point of failure. A mean contact width of ~50 nm is estimated by SEM measurements across both channels, with an estimated ~5-10% deviation along the device length. Post-breakdown imaging suggests failure at the Ti/Au contacts (red arrow). Measured current densities (33-37 $MA/cm^2$ at 80 K ambient) are quite similar to those reported in Figure 4d for wide $WTe_2$ devices. This is *unlike* short graphene nanoribbons (GNRs) which show *higher* current density than large graphene devices due to their larger thermal healing length which facilitates heat sinking to the contacts (for GNRs).[3] This highlights that the current density limitation of short WNRs is intrinsic, due to their low in-plane thermal conductivity.